\documentclass[english]{IEEEtran}
\usepackage[T1]{fontenc}
\usepackage[latin9]{inputenc}
\usepackage{amsmath}
\usepackage{graphicx}
\usepackage{babel}
\begin{document}
\title{Quantum Divide and Compute: Hardware Demonstrations and Noisy Simulations}
\author{Thomas Ayral,\IEEEauthorrefmark{1}\IEEEauthorrefmark{4} Fran\c{c}ois-Marie
Le R\'egent,\IEEEauthorrefmark{1}\IEEEauthorrefmark{2}\IEEEauthorrefmark{3}\IEEEauthorrefmark{4}
 Zain Saleem,\IEEEauthorrefmark{3}\IEEEauthorrefmark{4} Yuri Alexeev,\IEEEauthorrefmark{3}
Martin Suchara\IEEEauthorrefmark{3}\\\IEEEauthorrefmark{1}Atos Quantum Laboratory, Les Clayes-sous-Bois, France, \IEEEauthorrefmark{2}Ecole Polytechnique, Palaiseau, France,  \IEEEauthorrefmark{3}Argonne National Laboratory, Lemont, Illinois, United States of America\\\IEEEauthorrefmark{4}Equal contributions}
\maketitle
\begin{abstract}
Noisy, intermediate-scale quantum computers come with intrinsic limitations
in terms of the number of qubits (circuit ``width'') and decoherence
time (circuit ``depth'') they can have. Here, for the first time, we demonstrate
a recently introduced method that breaks a circuit into smaller
subcircuits or fragments, and thus makes it possible to run circuits that
are either too wide or too deep for a given quantum processor. We
investigate the behavior of the method on one of IBM's 20-qubit superconducting
quantum processors with various numbers of qubits and fragments.
We build noise models that capture decoherence, readout error, and gate imperfections for this particular processor. We then
carry out noisy simulations of the method in order to account for
the observed experimental results. We find an agreement within 20\%
between the experimental and the simulated success probabilities,
and we observe that recombining noisy fragments yields overall results
that can outperform the results without fragmentation.
\end{abstract}

\bstctlcite{IEEEexample:BSTcontrol}

Because of rapid technological progress, quantum processors of increasing
quality and size are becoming available, whether of the superconducting
\cite{Kjaergaard2020} or of the trapped-ion \cite{Bruzewicz2019}
type. Despite this steady improvement, these noisy, intermediate-scale
quantum (NISQ \cite{Preskill2018}) devices are still limited in both their number
of qubits (with, e.g., 53 qubits \cite{Arute2019}) and their coherence time.
Both constraints prevent one from performing quantum algorithms that
require a large number of qubits or operations. Peng \emph{et al.}
\cite{Peng2019} recently proposed a method to circumvent this limitation.
Basing their method on tensor-network techniques, they showed how
to decompose a circuit with a large quantum volume \cite{Cross2019}
into smaller subcircuits with quantum volumes compatible with NISQ
devices. 

Here, we show the first practical implementation of this method on
an actual 20-qubit quantum device for a Greenberger-Horne-Zeilinger
(GHZ) type of test circuit with a qubit count of up to 24 and various
fragments sizes. Rather than focusing on large qubit counts, we investigate
the extent to which this method can deal with decoherence in smaller
circuits through experimental runs and noisy simulation of this decoherence.
To this aim, we establish a precise noise model of IBM's 20-qubit Johannesburg
processor using available calibration data, and we use the model to simulate the experimental
results. This noisy simulation allows us to quantify and explain the
experimental results we obtain, and it paves the way to a noise-aware
optimization of this fragmentation technique.

\section{Methods: circuit fragmentation and noise modeling}

\subsection{Basics of circuit fragmentation}

\begin{figure}
\begin{centering}
\includegraphics[width=1\columnwidth]{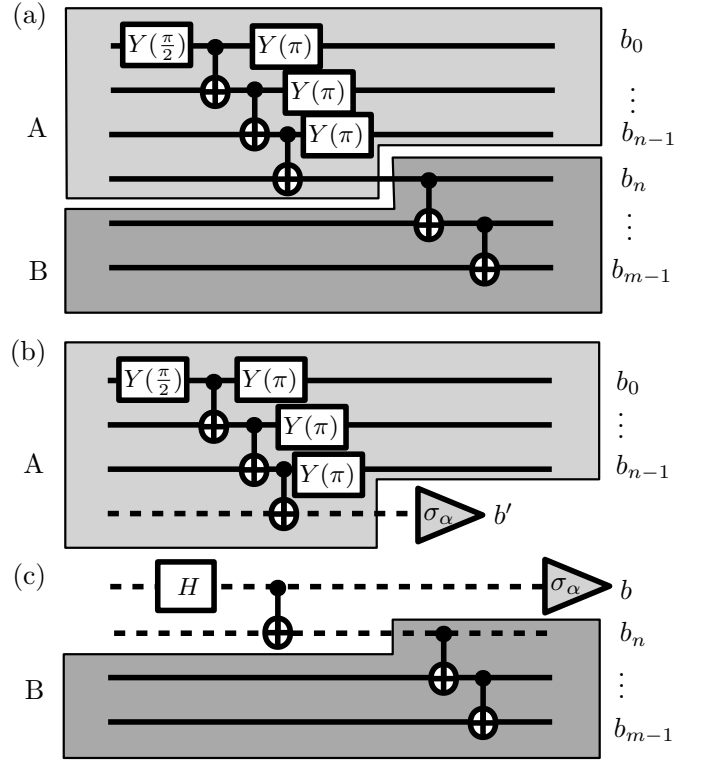}
\par\end{centering}
\caption{Fragmenting procedure for a $m=6$-qubit circuit. Qubit
with index $n$ is cut after the first controlled-NOT (CNOT) gate. Panels (b) and (c)
show the resulting two fragments. \label{fig:Sketch-fragmenting}}
\end{figure}

The execution of a quantum circuit on an $m$-qubit quantum computer
yields measurements in the form of bitstrings $\{(b_{0}\dots b_{m-1}),b_{i}\in\{0,1\}\}$
whose probability is given by Born's rule, $p(b_{0},\dots b_{m-1})=|\langle b_{0},\dots,b_{m-1}|U|\psi_{0}\rangle|^{2}$,
where $|\psi_{0}\rangle$ is the initial state of the quantum register
(here $|0\rangle^{\otimes m}$) and $U$ is the unitary operation
defining the quantum circuit. $U$ is composed of a sequence of local
unitary operations called quantum gates that can be represented as
the vertices of a graph. If the underlying graph can be broken into
disconnected components or ``fragments'' upon removal of edges,
the circuit's probability distribution $p(b_{0},\dots b_{m-1})$ can
be computed from the suitably modified probability distributions of
the fragments \cite{Peng2019}. For instance, the circuit in Fig.
\ref{fig:Sketch-fragmenting}(a) is represented by a graph that separates
into two disconnected components (light gray {[}A{]} and dark gray
{[}B{]}) when removing a single edge (here on qubit with index $n$
between the two CNOT gates). In this configuration, the full probability
distribution can be computed as
\begin{align}
 & p(b_{0}\dots b_{m-1})=\label{eq:final_formula_Bell}\\
 & \sum_{\alpha=X,Y,Z}\sum_{bb'\in\{0,1\}^{2}}\gamma_{\alpha}^{bb'}p_{A}^{\alpha}(b_{0}\dots b_{n-1};b')p_{B}^{\alpha}(b;b_{n}\dots b_{m-1})\nonumber 
\end{align}

with $\gamma_{X}^{bb'}=2\delta_{bb'}-1$, $\gamma_{Y}^{bb'}=-\gamma_{X}^{bb'}$
and $\gamma_{Z}^{bb'}=2\delta_{bb'}$. Here, $p_{A}^{\alpha}(b_{0}\dots b_{n-1};b')$
denotes the probability of measuring the bitstring $(b_{0}\dots b_{n-1},b')$
when measuring the final state of fragment $A$ along axis $\alpha$
for qubit $n$ (Fig. \ref{fig:Sketch-fragmenting}(b)), while $p_{B}^{\alpha}(b;b_{n}\dots b_{m-1})$
is the probability of getting bitstring $(b,b_{n}\dots b_{m-1})$
after preparing the first two qubits $(q,q_{n})$ (the first two qubits
of fragment $B$) in the $\left(|00\rangle+|11\rangle\right)/\sqrt{2}$
Bell state and measuring the final state of fragment $B$ with the
ancilla qubit measured along axis $\alpha$ (Fig. \ref{fig:Sketch-fragmenting}(c)).
This procedure can be repeated recursively to break the circuit into
ever smaller fragments.

With this procedure, a wide and deep quantum circuit can be fragmented
into smaller circuits that can be run on a NISQ processor. However,
doing so comes at a cost, in terms of the number of individual subcircuits to
be run, that is exponential in the number of removed edges or ``cuts''
\cite{Peng2019}.

In this work, we focus on the GHZ-type circuit shown in Fig. \ref{fig:Sketch-fragmenting}(a).
The resulting maximally entangled state, $\left(|0\rangle^{\otimes m/2}|1\rangle^{\otimes m/2}+(-)^{(m/2)\%2}|1\rangle^{\otimes m/2}|0\rangle^{\otimes m/2}\right)/\sqrt{2}$,
is very sensitive to decoherence and is therefore a good test case
for investigating the resilience of the method on noisy processors.

\subsection{Noise modeling and simulation}

To simulate the behavior of the method on noisy processors, we model
the processor errors by combining three error sources: decoherence
of the amplitude damping and dephasing types during qubit idling (inactive)
periods, readout errors, and gate imperfections. 

We set the amplitude damping, dephasing, and readout errors using calibration
data supplied on the IBM Quantum Experience platform. Averaging over
the 20 qubits of the chip, we find $T_{1}=65\mathrm{\mu s}$, $T_{2}=70\mathrm{\mu s}$,
and a readout error rate of $\gamma=4.1\%$. The $T_{1}$ and $T_{2}$
processes are modeled by the combination of the amplitude damping
(AD) and pure dephasing (PD) quantum channels defined by the Kraus
operators
\begin{align*}
\boldsymbol{K}_{0}^{\mathrm{A.D}} & =\left[\begin{array}{cc}
1 & 0\\
0 & \sqrt{1-p_{\tau_{\mathrm{idle}}}^{\mathrm{A.D}}}
\end{array}\right],\boldsymbol{K}_{1}^{\mathrm{A.D}}=\left[\begin{array}{cc}
0 & \sqrt{p_{\tau_{\mathrm{idle}}}^{\mathrm{A.D}}}\\
0 & 0
\end{array}\right],\\
\boldsymbol{K}_{0}^{\mathrm{P.D}} & =\left[\begin{array}{cc}
1 & 0\\
0 & \sqrt{1-p_{\tau_{\mathrm{idle}}}^{\mathrm{P.D}}}
\end{array}\right],\boldsymbol{K}_{1}^{\mathrm{P.D}}=\left[\begin{array}{cc}
0 & 0\\
0 & \sqrt{p_{\tau_{\mathrm{idle}}}^{\mathrm{P.D}}}
\end{array}\right],
\end{align*}

where $\tau_{\mathrm{idle}}$ is the duration of the idling period
during which the noise acts, $p_{\tau_{\mathrm{idle}}}^{\mathrm{A.D}}=1-e^{-\tau_{\mathrm{idle}}/T_{1}}$and
$p_{\tau_{\mathrm{idle}}}^{\mathrm{P.D}}=1-e^{-2\tau_{\mathrm{idle}}/T_{\varphi}}$,
with $\frac{1}{T_{\varphi}}=\frac{1}{T_{2}}-\frac{1}{2T_{1}}$. To
determine the idling durations, we assume the following durations
for the gates: 200 ns for the CNOT gate, and 20 ns for the single-qubit
gates. As for the readout errors, we choose to model them as a single-qubit
relaxation (amplitude damping) process during the measurement time.
The corresponding 2-outcome positive-operator valued measure (POVM)
has elements $\{\boldsymbol{E},\boldsymbol{I}-\boldsymbol{E}\}$,
with
\[
\boldsymbol{E}=\left(\begin{array}{cc}
0 & 0\\
0 & 1-\gamma
\end{array}\right),
\]
where $\gamma=1-e^{-t_{\mathrm{meas}}/T_{1}}$. We check that the
measurement duration $t_{\mathrm{meas}}$ we infer from the experimental
calibration error rate $\gamma$, namely $t_{\mathrm{meas}}=2.75\mu\mathrm{s}$,
is consistent with usual values for this duration.

We model the gate imperfections using a simple depolarizing noise
channel following each one-qubit gate, with Kraus operators
\begin{align*}
\boldsymbol{K}_{0}^{D} & =\sqrt{1-p_{(1)}^{D}}\boldsymbol{I},\\
\boldsymbol{K}_{i}^{D} & =\sqrt{p_{(1)}^{D}}\boldsymbol{\sigma}_{i},\;\;i=1,2,3
\end{align*}

where $\boldsymbol{\sigma}_{i}$ denote the Pauli spin matrices. For
the two-qubit (CNOT) gates, we use the tensor product of the above
depolarizing channel to mimic two-qubit errors after each CNOT gate.
We adjust the depolarizing probabilities $p_{(1)}^{D}$ and $p_{(2)}^{D}$
to have the error channels match given average process fidelities
$\mathcal{F}_{\mathrm{avg}}^{(1)}$ and $\mathcal{F}_{\mathrm{avg}}^{(2)}$
(as defined in e.g \cite{Gilchrist2005}) or equivalently average
errors $\mathcal{\epsilon}_{\mathrm{avg}}^{(1)}$ and $\mathcal{\epsilon}_{\mathrm{avg}}^{(2)}$
(with $\mathcal{F}_{\mathrm{avg}}=1-\epsilon_{\mathrm{avg}}$). $\mathcal{\epsilon}_{\mathrm{avg}}^{(1)}$
and $\mathcal{\epsilon}_{\mathrm{avg}}^{(2)}$ are themselves fixed
using the qubit-averaged calibration error rates supplied by IBM Quantum
Experience, $\mathcal{\epsilon}_{\mathrm{avg}}^{(1)}=0.041\%$ and
$\mathcal{\epsilon}_{\mathrm{avg}}^{(2)}=0.202\%$.

We use the obtained Kraus operators to simulate the noisy evolution
combined with fragmentation. Prior to the noisy simulation, the circuit
is compiled to comply with the target processor's qubit connectivity
graph using the Atos Quantum Learning Machine (QLM)'s dedicated \emph{Nnizer}
plugin. This results in longer circuits owing to the (optimized) insertion
of SWAP gates whenever needed. The noisy simulations are carried out
on the QLM using density-matrix-based simulations.

\section{Results}

\begin{figure}
\begin{centering}
\includegraphics[width=1\columnwidth]{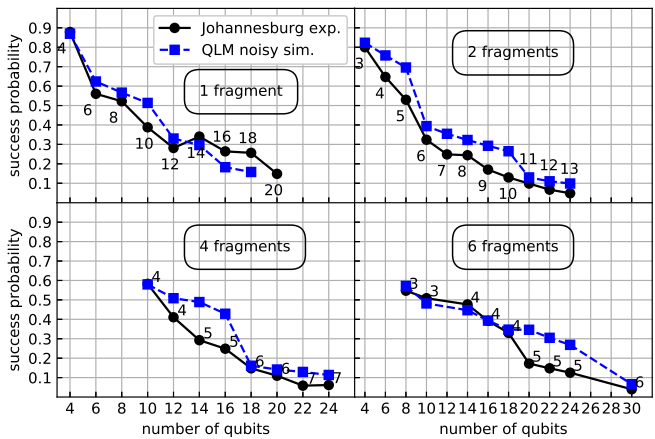}
\par\end{centering}
\caption{Success probability as a function of circuit size (number of qubits)
for various numbers of fragments using IBM's Johannesburg processor
(circles and solid black lines) and Atos QLM noisy simulation (squares
and dashed blue lines). The black numbers next to the black circles
indicate the maximum fragment size (in number of qubits) for the given
number of fragments and qubits.\label{fig:Success-probabillity-IBM-vs-QLM}}
\end{figure}

We implemented the circuit fragmentation procedure and tested it on
an experimental qubit platform, IBM's 20-qubit Johannesburg processor,
comprising superconducting transmon qubits arranged in a two-dimensional
grid. We accessed this processor via the IBM Quantum Experience cloud
platform and used the Qiskit programming framework to describe the
circuits. As a proxy for the quality of the final result, we calculated
the following sum of probabilities
\begin{equation}
P_{\mathrm{success}}\equiv p\left(|0\rangle^{\otimes m/2}|1\rangle^{\otimes m/2}\right)+p\left(|1\rangle^{\otimes m/2}|0\rangle^{\otimes m/2}\right),\label{eq:p_success_def}
\end{equation}
which is unity in the absence of any noise.

The experimental and noisy simulation results for up to 30 qubits
are shown in Fig. \ref{fig:Success-probabillity-IBM-vs-QLM}.
This figure includes the statistical error bars (standard error of the mean) on the probabilities after recombination. These errors originate from the finite number of shots (8192) per fragment. We computed them using resampling. Because of the large number of shots, they are comprised within the size of the datapoints and therefore do not appear on the graph.

The
one-fragment case (top-left panel), corresponding to running the original
circuit without fragmentation, will serve as our reference curve.
It displays a marked decrease in the success probability as the number
of qubits increases.
For all fragment numbers, the values obtained for the success probability obtained experimentally and with noisy simulation
agree within 20\% (in absolute values).
In particular, discontinuities and even some of the sign changes of
the slope of $P_{\mathrm{success}}$ are captured by noisy simulations.
The drops in success probability in going from a fragment size of
5 to a fragment size of 6 (and similarly 10 to 11 and 15 to 16) are
easily accounted for by the topology requirements of the chip (in
the absence of qubit relabeling, running a fragment of size 6 will
require introducing SWAP gates to perform a CNOT gate between qubits
of indices 4 and 5, which are not nearest neighbors on the chip).
The noisy simulations tend to overestimate the success probability compared to the experimental results. Uncaptured phenomena such as temporal and spatial (crosstalk)
noise likely account for the discrepancy. 

Remarkably, both experimental and noisy simulation results show that
increasing the number of fragments allows us to reach reasonable success
probabilities as the circuit sizes increase: thus, the success rate
drop after 4 qubits for the one-fragment case only occurs for circuit
sizes of 8 and 16 qubits when breaking the circuit into 2 and 4 fragments,
respectively (for the 6-fragment case, the experimental values show
a drop after 18 qubits, while the noisy simulation show the same drop
after 24 qubits). Thus, the method makes it possible not only to perform computations
for circuit sizes exceeding the chip's size (see, e.g, the $m=22,24,30$
runs), but also to obtain better success probabilities for smaller
circuit sizes.

\section*{Acknowledgment}

This research used resources of the Oak Ridge Leadership Computing
Facility, which is a DOE Office of Science User Facility supported
under Contract DE-AC05-00OR22725. This research also used the resources
of the Argonne Leadership Computing Facility, which is DOE Office
of Science User Facility supported under Contract DE-AC02-06CH11357.
Yuri Alexeev, Zain H. Saleem, and Martin Suchara were supported by
the DOE, Office of Science, under Contract DE-AC02-06CH11357. The
compilation and noisy simulations were performed using Argonne National
Laboratory's and Atos Quantum Laboratory's Quantum Learning Machines.

\bibliographystyle{IEEEtran}
%\bibliography{cutting_proj}
\bibliography{fragmenting_draft}

\end{document}